\documentclass[fleqn,10pt]{wlscirep}
\usepackage[utf8]{inputenc}
\usepackage[T1]{fontenc}
\usepackage{bm}
\usepackage{graphicx}
\usepackage{wrapfig}

\title{Photocurrent as a multi-physics diagnostic of quantum materials}

\author[1,2]{Qiong Ma$^{*,}$}
\author[3]{Roshan Krishna Kumar}
\author[4]{Su-Yang Xu}
\author[3,5]{Frank H.L. Koppens}
\author[6]{Justin C.W. Song$^{\dag,}$}
\affil[1]{Department of Physics, Boston College, Chestnut Hill, MA, USA}
\affil[2]{Canadian Institute for Advanced Research, Toronto, Canada}
\affil[3]{ICFO-Institut de Ciencies Fotoniques, The Barcelona Institute of Science and Technology, 08860 Castelldefels (Barcelona), Spain}
\affil[4]{Department of Chemistry and Chemical Biology, Harvard University, Cambridge, MA, USA}
\affil[5]{ICREA – Institució Catalana de Recerça i Estudis Avancats, 08010 Barcelona, Spain}
\affil[6]{Division of Physics and Applied Physics, School of Physical and Mathematical Sciences, Nanyang Technological University, Singapore 637371}
\affil[*]{e-mail: qiong.ma@bc.edu}
\affil[$\dag$]{e-mail: justinsong@ntu.edu.sg}

\begin{document}

\begin{abstract}
The photoexcitation life-cycle from incident photon (and creation of photoexcited electron hole pair) to ultimate extraction of electrical current is a complex multi-physics process spanning across a range of spatio-temporal scales of quantum materials. While often viewed through a device-technology lens, photocurrent is a key observable of the life-cycle that is sensitive to a myriad of physical processes across these scales. As a result, photocurrent is emerging as a versatile probe of electronic states, Bloch band quantum geometry, quantum kinetic processes, and device characteristics of quantum materials. This review outlines the key multi-physics principles of photocurrent diagnostics. In particular, we describe how the fundamental link between light-matter interaction and quantum geometry renders photocurrent a wavefunction-sensitive probe capable of resolving bandstructure and characterizing topological materials. We further highlight the sensitivity of the photoexcitation life-cycle to relaxational processes which in turn enables photocurrent to disentangle distinct types of carrier scattering that can range from femtosecond to nanosecond timescales. We survey the intrinsically nonlocal character of photocurrent collection that allows new types remote sensing protocols and photocurrent nanoscopy. These distinctive features underscore photocurrent diagnostics as a novel multi-physics probe for a growing class of quantum materials with properties governed by physics spanning multiple spatio-temporal scales. 
\end{abstract}

\flushbottom
\maketitle

\thispagestyle{empty}

\noindent \textbf{Key points:} 
\begin{itemize}
    \item The transduction of light into electrical signals (photocurrent) in quantum materials involves physical phenomena across multiple spatio-temporal scales and, therefore, photocurrent stands out as a multi-physics diagnostic tool of quantum materials. 
    \item The long-range collection of locally generated photocurrent, as mediated by the streamlines of diffusion currents, enables a ``remote" sensing of local symmetry breaking such as p-n junctions and edges. 
    \item Photocurrent spectroscopy can probe charge, spin, and collective excitations with enhanced signal-to-noise characteristics and resolution for atomically-thin materials with low optical weight.
    \item The intimate connection between light-matter interaction and Bloch band quantum geometry renders bulk geometric photocurrent highly sensitive to the crystal symmetry and light polarization. 
    \item Technique innovation including near-field photocurrent probes as well as ultrafast photocurrents grant high spatio-temporal resolution, providing a range of new photocurrent diagnostics tools for quantum materials.
\end{itemize}

\section*{Introduction}
The transduction of light (charge-neutral photons) into electrical signals (charged electrons) is one of the most fundamental processes in electronic materials and devices.~\cite{nelson2003physics} This transduction involves a complex photoexcitation lifecycle process that involves a menagerie of physical phenomena across multiple spatio-temporal scales. Photocurrent stands out as a key observable that is active throughout this lifecycle enabling it as an effective sensor of how a quantum material is pushed out-of-equilibrium by light.\\
 
Due to its multi-physics and multi-scale nature, extracting detailed information about quantum materials from photocurrent is often simultaneously challenging but rewarding. Insight into the detailed lifecycle of photocurrent has recently become forthcoming due to advances in technology, technique, and theory. For instance, experimental technique innovation (e.g., time-resolved photocurrent~\cite{sun2012ultrafast,graham2013photocurrent,tielrooij2015generation} and scanning nearfield photocurrent nanoscopy~\cite{woessner2016near, lundeberg2017thermoelectric, sunku2021hyperbolic, hesp2021nano}), theoretical advances (e.g., quantum geometrical origin of bulk photocurrents~\cite{Morimoto:2016iu}), the discovery and engineering of high quality quantum materials (e.g controlling photocurrents in van der Waals heterostructures~\cite{akamatsu2021van, ma2022intelligent}) have not only enabled an enhanced understanding of the lifecycle, but have exploited photocurrent as a diagnostic tool.\\

\begin{figure}[ht]
\centering
\includegraphics[scale=0.5]{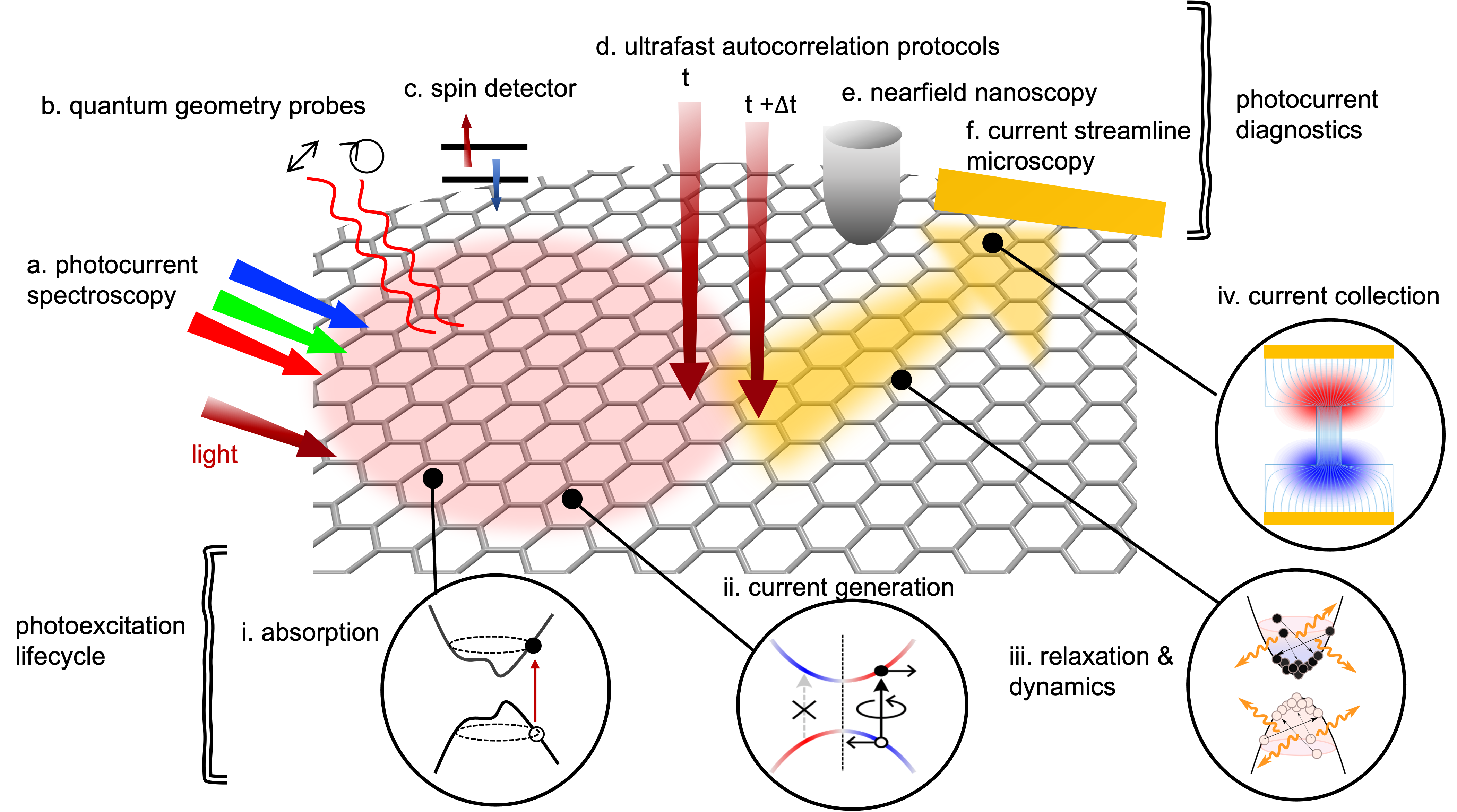}
\caption{{\bf Multi-physics photocurrent diagnostics enables to probe each stage of the photoexcitation lifecycle}. Photoexcitation lifecycle (i-iv, bottom) consists of broad stages that span multiple spatio-temporal scales. Probing photocurrent during each stage of the lifecycle  provides distinct snapshots of the electronic state, phase, quantum kinetic processes, and device characteristics. These snapshots are enabled by novel photocurrent diagnostics (for example, a-f, top) that are highly sensitive to electronic degrees of freedom.} 
\label{fig_overview}
\end{figure}

In order to track the photoexcitation lifecycle, it is useful to consider the various stages of the photoexcitation cascade of an electronic system from initial photoexcitation to carrier relaxation and evolution back to equilibrium. Photocurrent at each stage of this multi-scale process can provide snapshots of the electronic quantum state, phase, as well as its symmetries. When applied to large-scale material samples/devices, it can also deliver insight into device characteristics. There are four broad stages to the lifecycle (as illustrated in Fig.~\ref{fig_overview}): (i) photoexcitation and absorption (ii) local current generation, (iii) relaxation and carrier dynamics, and (iv) propagation and collection. (i) and (ii) occur at the fastest timescales (several to hundreds of femtoseconds), whereas the timescales characterizing (iii) and (iv) often depend on material specific properties and 
span a wide set of timescales that range from hundreds of femtoseconds to tens of picoseconds, and, in some materials, even longer time scales.~\cite{dong2015electron} We note that these processes [e.g., (iii) and (iv)] can in principle occur in parallel enabling rich and complex photocurrent dynamics.\\

The properties of photocurrent during each of these stages contain important information of the electronic state and, as a result, can be exploited as operational principles for novel photocurrent diagnostics of quantum materials. Take for instance stage (iv). Photocurrent can possess a highly spatially nonlocal and long-range character. This long-range character enables a ``remote'' sensing of local photo-induced currents generated far away from current collecting contacts and new principles for a range of photocurrent ``probe locally'' and ``sense remotely'' spatial diagnostics (see e.g., Fig.~\ref{fig_overview}{\bf e} and {\bf f}). Similarly, the sensitivity to the degree in which an electronic distribution is pushed out-of-equilibrium in stage (iii) grants ultrafast photocurrents the ability to disentangle quantum kinetic and relaxational/scattering processes. Ultrafast autocorrelation techniques (Fig.~\ref{fig_overview}{\bf d}) and time-domain photocurrent effectively exploit this distinct sensitivity. The intimate connection between light-matter interaction and Bloch band quantum geometry in stage (i) and (ii), e.g., through bulk photocurrents without a p-n junction, 
renders photocurrent an ideal probe of band geometry as well as the intricate low-energy features of correlated bandstructure~\cite{ma2021topology,orenstein2021topology} (see.e.g, spectroscopic and polarization dependent photocurrent techniques Fig.~\ref{fig_overview}{\bf a},{\bf b}).\\

In this review, we highlight the multi-scale and multi-physics life-cycle of photocurrent in quantum materials, with a particular focus on the key principles of how photocurrent at each stage of the life-cycle can be exploited as a sensitive probe of intricate lives of quantum materials when pushed out-of-equilibrium. 

\section*{Photocurrent collection: spatial diagnostics and local probes}

\begin{figure}[t]
\centering
\includegraphics[scale=1.0]{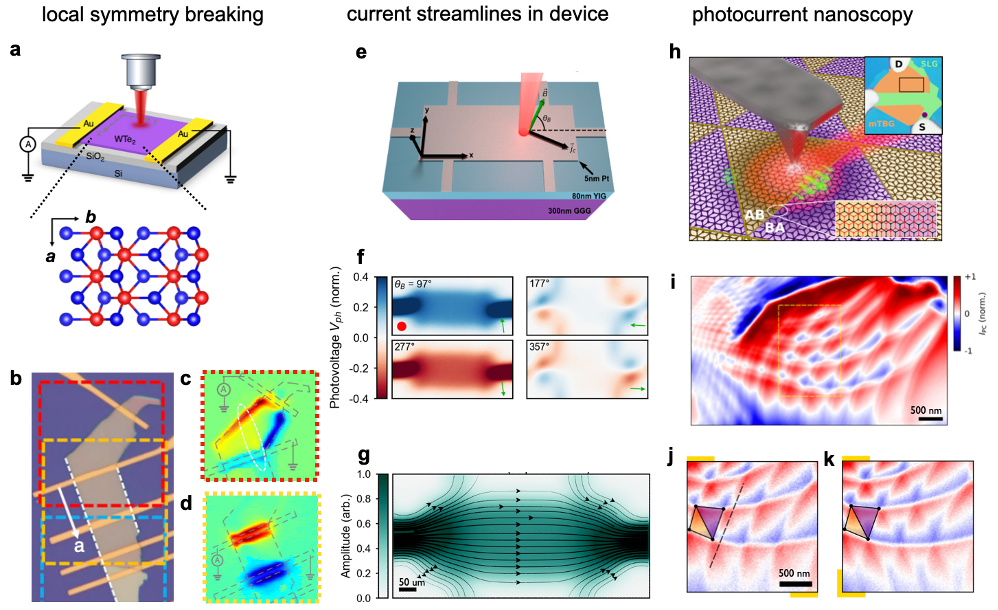}
\caption{\textbf{Non-local sensing of photocurrent enables to image the spatial characteristics of quantum materials and devices:} (a,b) Imaging of local symmetry breaking in a quantum material via photocurrent. (a) scanning photocurrent set-up wherein a focused laser beam excites a local photocurrent density in a quantum material (in this case, WTe$_2$) and global photocurrent is collected at remote contacts (yellow). (inset) crystal structure of WTe$_2$ exhibiting a mirror axis perpendicular to a. (b) Image of WTe$_2$ quantum device with current collecting contacts shown in yellow. (c) along edges that break mirror symmetry, photocurrent propagating along the edge can be remotely sensed at the current collecting contacts. In contrast, (d) along edges parallel to a, mirror symmetry enforces a vanishing photocurrent parallel to the edge (photocurrent is only visible close to contacts). Adapted from Ref.~\citenum{wang2019robust}. (e-g) Photocurrent can enable direct imaging of current streamlines in a device. (e) By employing a focused laser beam spot as well as an in-plane magnetic field in a magnetic heterostructure Pt/YIG, the position as well as the direction of the local photocurrent density can be controlled by the laser beam and the magnetic field direction respectively. (f) Spatial maps of photovoltage display different patterns as the direction of magnetic field is rotated. By tracking the magnetic field direction that maximizes the collected photovoltage as a function laser spot position (as well as its amplitude of photovoltage), current steamlines in a device can be naturally mapped out. Adapted from Ref.~\citenum{mayes2020current}. (h-k) In combination with a (h) scanning near field tip, “photocurrent nanoscopy” enables to image sub-micron (“nano-scale”) spatial features in quantum materials. In such a set-up, far field light is incident on the quantum material but is concentrated on a highly localized spot using a nearfield tip. (inset) Photocurrent is collected at remote contacts marked “S” and “D”. In low-twist angle twisted bilayer graphene (TBG), (i) photocurrent nanoscopy enables to image the AB/BA domain structure. (j,k) Changing the specific current collecting contacts wherein the global photocurrent is collected (see yellow) produce subtle changes in the photocurrent spatial map characteristic of the changing Shockley-Ramo streamlines. Adapted from Ref.~\citenum{hesp2021nano}.}
\label{fig_nonlocal}
\end{figure}

A particularly attractive feature of photocurrent is its long-range and spatially non-local character. A spatially local photo-induced current (local photocurrent) generated in a tight laser beam spot can often be sensed far away at remote current collecting/voltage sensing contacts, see e.g. Fig.~\ref{fig_nonlocal}\textbf{a}. Strikingly, this remote sensing can occur over distances far larger than the electron mean-free path of the device.~\cite{park2009imaging, cao2016photo, ma2019giant, wang2019robust} Its long-range character is a universal characteristic of many opto-electronic devices and is mediated by ambient carriers in an electronic material. Similar to the way the voltage on a parallel-plate capacitor can be sensitive to the local charge distribution between the plates, the generation of a local photocurrent induces an electromotive force that drive long-range diffusion currents into remote device contacts.~\cite{song2014shockley}\\

The long-range photocurrent behavior is often accounted for via the “Shockley-Ramo theorem” ~\cite{shockley1938currents,ramo1939currents,song2014shockley} that describes device specific streamline patterns along which diffusion currents prefer to flow (see Box 1 for description and example streamline patterns). Since ambient carriers transmit the signal to remote contacts, the streamlines are agnostic to the precise mechanism of local photocurrent generation. When local photocurrent (a vector quantity) is aligned (anti-aligned) with the streamlines, they produce a maximal and positive (negative) response at the contacts; when local photocurrent is generated in a region with a high (small) density of streamlines, the global current collected at remote contacts is maximized (minimized). This simple framework has been broadly used to account for photocurrent signals in a variety of materials and geometries.~\cite{hesp2021nano, lundeberg2017thermoelectric, sunku2021hyperbolic,gabor2011hot,wang2019robust, ma2019giant, cao2016photo, woessner2016near, seifert2019quantized, shao2021nonlinear, mayes2020current}\\

{\bf Photocurrent spatial maps -- } Remote sensing of photocurrent enables a unique type of diagnostic of spatial features of quantum materials. A typical photocurrent microscopy set-up is illustrated in Fig.~\ref{fig_nonlocal}a where tight laser beam spot scans a device sample while the current (or the operationally equivalent photovoltage) is collected at remote contacts – the photocurrent as a function of the position of the laser beam spot are often termed photocurrent spatial maps and enable to image a myriad of spatial features in quantum materials.\cite{park2009imaging,xu2010photo,lemme2011gate,gabor2011hot,wang2019robust, seifert2018spin, seifert2019quantized, shao2021nonlinear,mayes2020current} 

Because local photocurrent is a vector, photocurrent spatial maps are particularly sensitive to local symmetry breaking.~\cite{ma2019giant,wang2019robust}. For instance, Ref.~\citenum{wang2019robust} used scanning photocurrent spatial maps to identify local mirror symmetry breaking along edges in a WTe$_2$ device: along edges where local crystalline mirror symmetry was broken (red boxes in Fig.~\ref{fig_nonlocal}\textbf{b} and \textbf{c}), large photocurrent was measured. In contrast, along edges where crystal mirror symmetry along the edge was preserved (orange box Fig.~\ref{fig_nonlocal}\textbf{b} and \textbf{d}), photocurrent along the edge vanished; the photocurrent in the latter maps of Fig.~\ref{fig_nonlocal}\textbf{d} only arose from the contacts. This demonstrates how photocurrent maps can be used to identify crystalline symmetry breaking in a quantum device that can be otherwise challenging to diagnose using purely optical microscopy techniques. The same technique can be used to directly probe the topological surface or edge states of a range of topological electronic materials.~\cite{seifert2018spin, seifert2019quantized, shao2021nonlinear, dantscher2017photogalvanic, plank2018review}\\

{\bf Streamline imaging -- } Since photocurrent measured at device contacts depend directly on the streamline patterns along which diffusion currents prefer to flow, photocurrent maps can be used as a means of directly imaging streamline spatial patterns in quantum devices. For example, in Ref.~\citenum{mayes2020current}’s magnetic heterostructure Pt/YIG, out-of-plane heating from a localized laser spot generates an in-plane local photocurrent whose in-plane direction can be completely controlled by a magnetic field, see Fig.~\ref{fig_nonlocal}\textbf{e}. In-situ control of local photocurrent density produces strikingly different photocurrent spatial maps for different magnetic field directions, Fig.~\ref{fig_nonlocal}\textbf{f}; these patterns can be in turn be used to map out the native current streamlines patterns of the device (in this case, a Hall bar), see Fig.~\ref{fig_nonlocal}\textbf{g} ~\cite{mayes2020current}, much as streamers are used in wind tunnels to indicate the direction and speed of airflow. Such streamlines are key characteristics of semiconductor devices – photocurrent maps enable to directly access them.\\

{\bf Nearfield enabled photocurrent nanoscopy and local probes -- } When combined with nearfield techniques, scanning photocurrent spatial maps can achieve dramatically improved resolution~\cite{woessner2016near, lundeberg2017thermoelectric, hesp2021nano, sunku2021hyperbolic}. 
Far-field irradiation incident on an atomic force microscopy probe 
(Fig.~\ref{fig_nonlocal}\textbf{h}) produces a near-field electromagnetic field at the tip apex that allows a highly localised photoexcitation at length scales of $10-30\,{\rm nm}$, and thus far below the diffraction limit. Scanning the tip over the sample while probing photocurrent through remote contacts (see inset Fig.~\ref{fig_nonlocal}\textbf{h}) can reveal spatially localized features of several tens to a hundred nanometers in size e.g., grain boundaries in chemical vapor deposited graphene~\cite{woessner2016near}, as well as domain walls in twisted bilayer graphene~\cite{hesp2021nano, sunku2021hyperbolic}, see Fig.~\ref{fig_nonlocal}\textbf{i}. Even as local photocurrent density is generated at highly localized domain walls, they are nevertheless sensitive to the global “Shockley-Ramo” streamlines that subtly change as the current collecting contacts are changed, see Fig.~\ref{fig_nonlocal}\textbf{j},\textbf{k}. Such a technique can be used as a high-throughput means of identifying other types of localized spatial features such as and puddles~\cite{woessner2016near}, and, crucially, linking them with device performance.

\section*{Photocurrent as a probe of electronic DOF: charge, spin, and collective excitations}

\begin{figure}[ht]
\centering
\includegraphics[width=\linewidth]{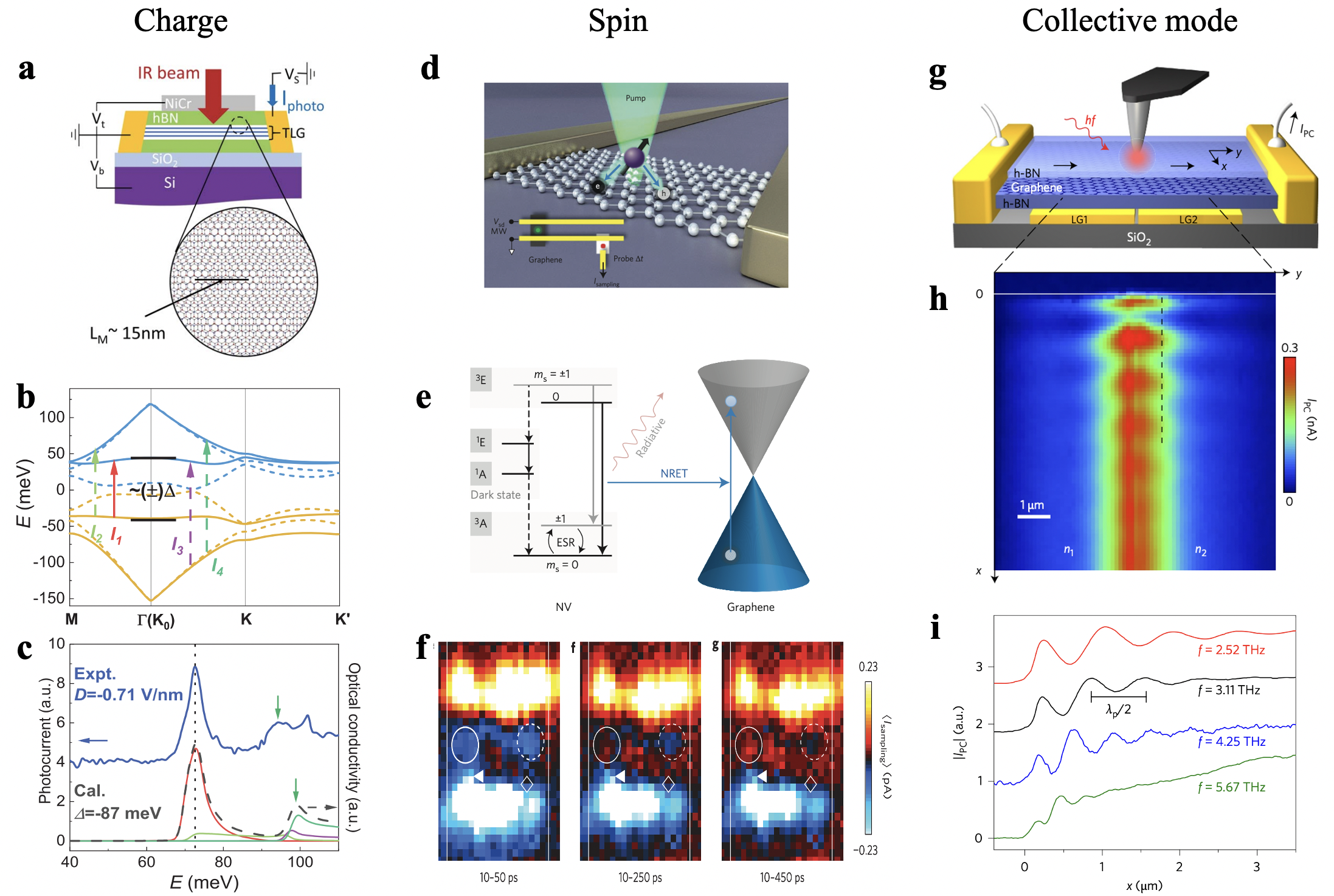}
\caption{\textbf{Photocurrent measurements of charge, spin, and collective excitations in quantum systems:} (a-c) Photocurrent infrared spectroscopy can probe the charge excitations in a correlated ABC trilayer graphene-boron nitride moiré system. Due to the low photon energies and high signal-to-noise ratio, photocurrent infrared spectroscopy can resolve tiny energy gaps between moiré minibands and also emergent correlated charge gaps. Adapted from Ref.~\citenum{yang2022spectroscopy}. (d-f) Detection of the dynamics of quantum spins in nitrogen vacancies through energy transfer into graphene probed by ultrafast photocurrent mapping. Adapted from Ref.~\citenum{brenneis2015ultrafast}. (g-i) Detection of THz propagating plasmons in graphene with a spatially-resolved photocurrent nanoscope. The nanotip is scanned across the entire sample and an electrical current is measured simultaneously. Local heating at the p-n junction caused by the THz nanotip generates a PTE current. The photocurrent is generated by a photothermoelectric effect due to the absorption of THz collective modes. Intriguingly, near the p-n junction, there are photocurrent oscillations that are perpendicular to the graphene edge. Those oscillations reveal the excitation of propagating surface plasmons. Adapted from Ref.~\citenum{alonso2017acoustic}.}
\label{fig_excitation}
\end{figure}

Because photocurrent relies on the transport of electronic degrees of freedom (DOF), it is uniquely sensitive to the charge, spin, and collective electronic characteristics of materials. Indeed, even though photocurrent relies on an initial optical excitation, its electrical readout (see e.g, Fig.~\ref{fig_excitation}\textbf{a}) enables to isolate electronic DOFs; as will be discussed below, this enables to overcome significant experimental and environmental challenges such as low optical weight to achieve enhanced signal-to-noise characteristics and resolution.\\

{\bf Photocurrent spectroscopy --} An important example is photocurrent spectroscopy (and its related photoconductivity spectroscopy), that is widely applied to measure the absorption spectrum of electronic materials. Both photocurrent and photoconductivity operate in a similar fashion with one key distinction: photoconductivity originates from light-induced changes to the electrical conductivity and requires an external electric bias to generate a measurable photocurrent. In photocurrent spectroscopy, various types of light sources can be used that include monochromatic sources with continuously tunable wavelengths~\cite{mak2010atomically} as well as white sources where multiple wavelengths are mixed.~\cite{ju2017tunable} The former utilizes high-power wavelength tunable sources, with photocurrent directly measured as a function of wavelength thereby generating a photocurrent spectrum~\cite{mak2010atomically}. Sources can take on ranges in the visible~\cite{mak2010atomically} and even the THz regimes~\cite{alonso2017acoustic}. In using white sources, e.g., in the infrared and THz spectroscopy ranges, a Fourier transform methodology can be adopted to obtain the photocurrent spectrum.~\cite{ju2017tunable, yang2022spectroscopy}\\ 

While traditionally applied in conventional semiconductors to detect large bandgaps and trapping states in the visible spectrum, photocurrent spectroscopy has recently been successfully employed %to great effect 
in quantum materials to detect the intricate details of electronic structure and correlated behavior at low energy scales. Particularly powerful is the Fourier transform infrared (FTIR) photocurrent spectroscopy technique that can realize sensitive infrared and THz spectroscopy for small samples of 2D heterostructures.~\cite{ju2017tunable,han2021accurate, yang2022spectroscopy} Such photocurrent probes can achieve a much higher signal-to-noise ratio (as compared with purely optical methods) since the electrical current readout does not pick up the environmental optical background.~\cite{yang2022spectroscopy} Indeed, a recent FTIR photocurrent spectroscopy applied the technique to trilayer graphene-BN moiré superlattices~\cite{yang2022spectroscopy} enabled to resolve tiny moiré miniband gaps as well as emergent correlated charge gaps (Fig.~\ref{fig_excitation}\textbf{b-c}).\\

{\bf Fermi surface photocurrent probes --} In addition to probing highly-excited states, photocurrent can also be sensitive to the details of the electronic Fermi surface, especially in two-dimensional metals and semimetals where photocurrent generation is often dominated by the photothermoelectric effect (PTE). A dramatic example is graphene where the small electronic heat capacity and weak electron-phonon coupling lead to significant heating of the electronic system after optical absorption.  Elevated electronic temperatures at a p-n junction can drive hot electrons and holes in opposite directions, generating a PTE current.~\cite{xu2010photo, song2011hot, lemme2011gate, gabor2011hot} Such photocurrent is highly sensitive to the Fermi surface. For instance, since the PTE is controlled by the Seebeck coefficient that in turn depends on the Fermi surface (both size and polarity), PC can be highly sensitive to the local chemical potential enabling to identify charge puddles~\cite{woessner2016near}, transitions between Landau levels~\cite{nazin2010visualization}, as well as time-reversal symmetry breaking through a Nernst effect.~\cite{cao2016photo} From a broader point of view, PTE can be used as a thermometer of electronic temperature~\cite{gabor2011hot,sun2012ultrafast,graham2013photocurrent,tielrooij2015generation} allowing to track the dynamics of out-of-equilibrium hot carrier distributions and their associated ultrafast scattering processes, as elaborated below.\\

{\bf Photocurrent as spin probes -- } Beyond charge, PC can additionally be sensitive to a range of other electronic DOFs. For instance, via an energy transfer process between graphene and other quantum systems, the PTE can be used to detect excitation and dynamics in other quantum systems, such as spins in quantum defects. As discussed by Brenneis et al.~\cite{brenneis2015ultrafast} in a setup consisting of nitrogen-vacancy quantum emitter close to graphene, after optical excitation of the NV centers, the relaxation occurs both through radiative channels (luminescence) and non-radiative energy transfer into graphene, which is spin-dependent for NV centers. The latter excites electron-hole pairs in graphene which can be converted into electrical signals directly or through decaying into hot carriers to generate a photothermoelectric current. By measuring the temporal dynamics of the photocurrent in the NV-graphene hybrid system, it was found that the photocurrent dynamics is shorter than the intrinsic NV lifetime but longer than the graphene carrier dynamics. This renders photocurrent a novel electrical readout method of quantum spin transitions. This technique can be turned on its head wherein NV center sensing can be used to detect localized photocurrent generation.~\cite{wang2022visualizing} In spin-orbit coupled systems, photocurrent can even be used as a means of controlling spin: circular-polarized light can induce spin polarized photocurrents.~\cite{ganichev2003spin,ivchenko2017spin, mciver2012control, ganichev2014interplay}\\

{\bf Photocurrent enabled electrical read out of collective modes -- } Another electronic DOF beyond charge are those of collective excitations. Electrical read-out signatures of collective excitations (such as plasmons) in quantum materials are often challenging~\cite{du2017highly}. PTE’s Fermi surface and hot carrier sensitivity render it an ideal tool for an all-electrical detection of plasmons~\cite{lundeberg2017thermoelectric, alonso2017acoustic}, e.g., infrared and THz plasmons in graphene (Fig.~\ref{fig_excitation}{\bf g}-{\bf i}). In these, irradiation is focused onto a nanotip to assist the near-field excitation of strongly confined plasmons in graphene. The plasmons propagate radially away from the tip, reﬂecting at edges, interfaces, and defects. Then, the self-interfered plasmon wave decays into electronic heat; the subsequent heat diffusion spreads the electronic heat to the pn-junction, where a voltage is generated. This method can be used to probe the spatial profile of graphene plasmons, as well as its dispersion (Fig.~\ref{fig_excitation}{\bf i}). Beyond all-electrical readout of infrared and terahertz graphene plasmons~\cite{lundeberg2017thermoelectric, alonso2017acoustic}, 
propagating phonon-polaritons can also be electrically detected~\cite{woessner2017electrical}. The power of the technique as a physical probe for quantum optical non-local effects was demonstrated in Ref.~\citenum{lundeberg2017tuning}, through the observation of propagating THz plasmons while slowing them down to the electronic Fermi velocity using a gate. Collective polariton modes can also be studied using far-field photocurrent techniques when using certain custom device architectures that are dominated by a polariton-assisted photoresponse~\cite{bandurin2018resonant,castilla2020plasmonic}\\

This strong sensitivity to electronic DOF can be further exploited to probe other collective modes that may have small optical weight. A case-in-point are interlayer exciton states in bilayer transition metal dichalcogenide heterojunctions; while photoluminescence (a purely optical technique) is often the technique of choice, small optical weights of interlayer excitons make probing them via optics challenging. Photocurrent spectroscopy (an optical excitation and electrical readout technique) is uniquely sensitive to electronic degrees of freedom enabling to sensitively probe interlayer excitons~\cite{barati2017hot,vialla2019tuning} as well as their fine exciton-phonon coupled “vibronic” structure.~\cite{barati2022vibronic}. Photocurrent can even be sensitive to collective excitations in a proximal substrate: substrate phonon excitations can be remotely sensed via photocurrent induced in a proximal layer~\cite{badioli2014phonon}.

\section*{Quantum geometric photocurrents: wavefunction diagnostics}

At first blush, Bloch band quantum geometry and light-matter interaction may seem unrelated. The former characterizes how Bloch electronic states evolve as a function of momentum~\cite{xiao2010berry, provost1980riemannian}, 
while the latter (and its concomitant photocurrents) are often described within an electric-dipole interaction.~\cite{Sipe00} Recently, however, 
theory~\cite{Morimoto:2016iu, de2017quantized,moore2010confinement,sodemann2015quantum} has emphasized their intimate relationship: the same dipole matrix elements that describe the transition amplitudes between initial and final states during photoexcitation also form the basis for interband quantum geometry linking quantum geometrical quantities such as the quantum metric~\cite{ahn2020low} and Berry curvature~\cite{xiao2010berry} to optoelectronics. Indeed, optical selection rules for photoexcitation often arise hand-in-hand with non-trivial quantum geometry. A 
prominent example is that of valley selective circular dichroism in gapped Dirac materials such as gapped bilayer graphene or transition metal dichalcogenides~\cite{yao2008valley}: the opposite signs of Berry curvature in $K$ and $K'$ valleys enables circularly polarized light to selectively photoexcite electrons in distinct valleys granting access to valley optoelectronics such as a photoinduced valley selective Hall conductivity~\cite{mak2014valley, yin2022tunable}.\\

{\bf Quantum geometry and bulk photocurrents without p-n junctions --} Quantum geometry can manifest in optoelectronic phenomena far beyond that of optical absorption.~\cite{ma2021topology, orenstein2021topology} A striking example is the generation of bulk photocurrents in non-centrosymmetric materials, the so-called bulk photovoltaic effect or BPVE.~\cite{Sipe00, tan2016shift, Morimoto:2016iu,fridkin2001bulk,cook2017design} Unlike conventional photovoltaics for which photocurrents are generated at macroscopic p-n junctions, in BPVE photocurrents are generated uniformly throughout the bulk of a material in the absence of p-n junctions. Such bulk photocurrents possess a direction that depend on a combination of factors that include the polarization, quantum geometry of the material, as well as the presence/absence of symmetries.\\

Depending on the precise microscopic mechanism, such bulk photocurrents are sensitive to different aspects of the Bloch band quantum geometry (see Box 2 for description of bulk photocurrent mechanisms). For instance, for bulk photocurrents arising from interband transitions, circular (linear) injection currents formed from asymmetries in the photoexcited carrier’s group velocity are sensitive to the interband Berry curvature (quantum metric)~\cite{ahn2020low, de2017quantized}; similarly, linear (circular) shift photocurrents formed from real space displacements of electrons as they are photoexcited depend on the shift vector~\cite{Sipe00, von1981theory}.  In non-centrosymmetric metals, bulk photocurrents can even flow when irradiated with light frequencies below the interband optical transition gap. Such metallic photocurrents depend on other quantum geometrical quantities such as the intraband Berry curvature dipole ~\cite{moore2010confinement,sodemann2015quantum} – a full account of these are the subject of current intense investigation.~\cite{de2020difference,matsyshyn2019nonlinear,gao2021intrinsic, watanabe2021chiral,shi2022berry,onishi2022photovoltaic} Even in centrosymmetric materials, bulk photocurrents of a quantum geometric nature can be induced by non-vertical transitions mediated by finite momentum of polaritons.~\cite{shi2021geometric,xiong2021unblocking}\\

\begin{figure}[ht]
\centering
\includegraphics[width=\linewidth]{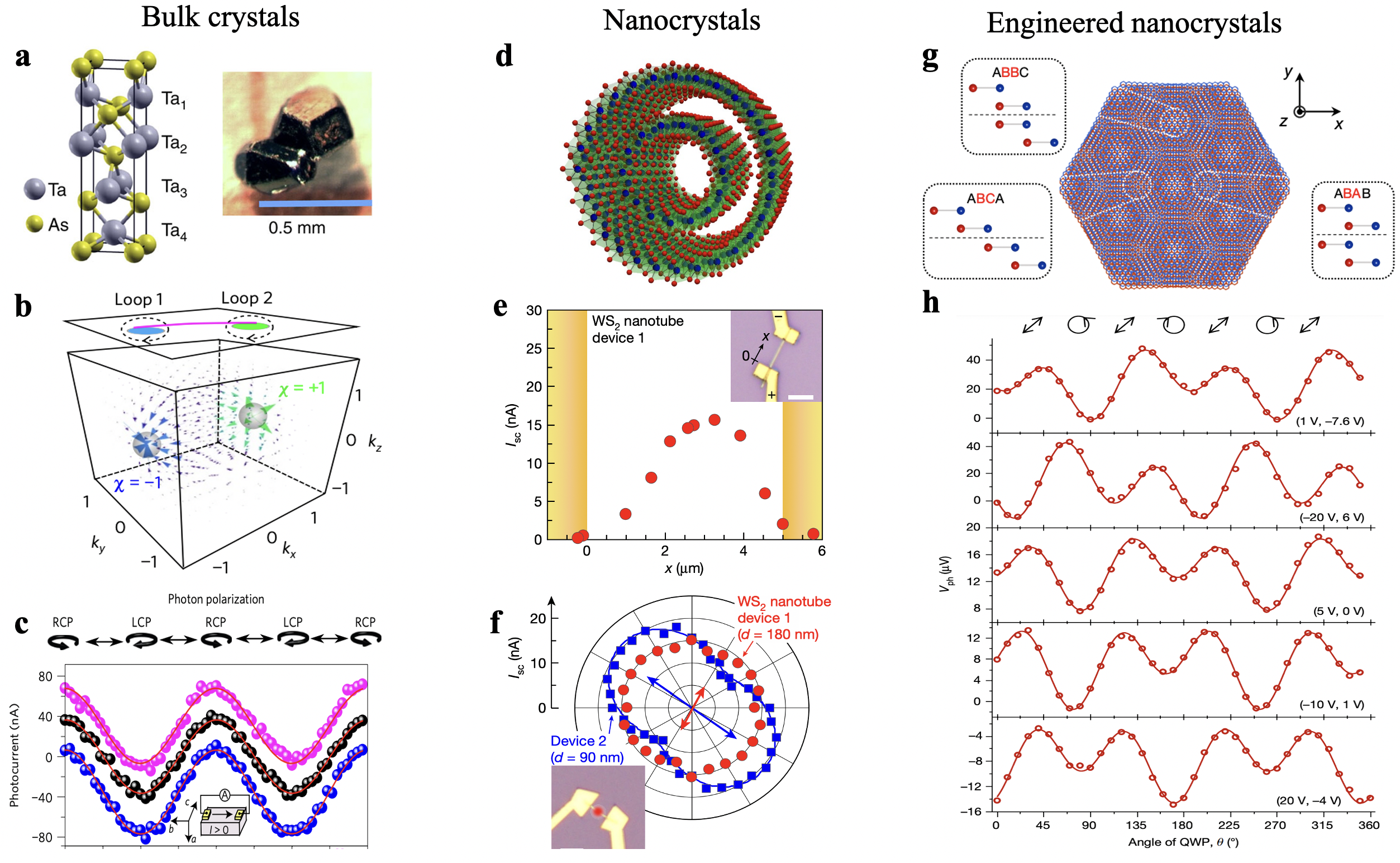}
\caption{\textbf{Quantum geometric photocurrents observed in a variety of materials with distinct dimensionalities and symmetry:} (a-c) Circular injection photocurrent arising from the chirality and Berry curvature of Weyl nodes in a  topological Weyl semimetal TaAs with polar structure. Panel a is the lattice structure and real picture of TaAs crystals, adpated from Ref.~\citenum{huang2015weyl}. Panels b-c are the $k$-space Berry curvature monopoles in a Weyl semimetal and photocurrent data of TaAs, adapted from Ref.~\citenum{ma2017direct}. (d-f) Crystal lattice and linear shift photocurrent in a WS$_2$ nanotube whose symmetry is lowered due to reduced dimensionality. Adapted from Ref.~\citenum{zhang2019enhanced}. (g-h) Lattice structure and photocurrents of twisted double bilayer graphene. The observation of highly tunable (by light polarization and external electric field) mid-infrared bulk photocurrent in an engineered two-dimensional heterostructure, twisted double bilayer graphene, arising from quantum geometry in the moiré bands. Adapted from Ref.~\citenum{ma2022intelligent}.}
\label{fig_geometry}
\end{figure}

{\bf Photocurrent diagnostics of quantum geometry -- } As a result of the intimate connection between bulk photocurrents and Bloch band quantum geometry, photocurrents have recently emerged as a powerful tool to study topological materials and emergent quantum phases in correlated electronic systems from the perspective of symmetry and quantum geometry. Low photon energies are often key to observing sizable quantum geometric photocurrents arising in topological materials or from correlated electrons. Large injection and shift currents were theoretically predicted for infrared and THz photons and experimentally reported by using mid-infrared photons in topological Weyl semimetals TaAs and TaIrTe$4$ (Fig.~\ref{fig_geometry}\textbf{a-c})~\cite{ma2017direct,osterhoudt2019colossal,ma2019nonlinear} and also a monolayer topological insulator WTe$_2$.~\cite{xu2018electrically}. A particularly interesting prediction is the quantized circular injection rate in Weyl semimetals with spatially chiral lattices where inversion and mirror symmetries are all broken.~\cite{de2017quantized,chang2018topological} Material candidates include CoSi and RhSi where quantization can be used as a means to empirically count the number of Weyl nodes filled. However, a precise measurement of the current injection rate is challenging due to the short momentum lifetime of photocarriers. In addition, theory found that the quantization of CPGE injection in chiral topological semimetals is generically removed by including perturbations from interactions~\cite{avdoshkin2020interactions}. Nonetheless, several experimental groups have led significant efforts in applying THz time-domain spectroscopy techniques with infrared pump to detect the rare quantization phenomenon in nonlinear responses.~\cite{rees2020helicity, ni2021giant} Such techniques are described in the following section on ultrafast photocurrent techniques.\\

{\bf Controlling quantum geometrical photocurrents in low dimensions -- } Given the sensitive dependence of quantum geometrical photocurrents on crystalline symmetries, low-dimensional materials have become a choice platform to control quantum geometrical photocurrents. By exploiting reduced crystalline symmetries in low-dimensional materials (e.g., in nanotubes) enhanced  bulk quantum geometrical photocurrents can be realized (Fig.~\ref{fig_geometry}\textbf{d-f}). The ability to stack two-dimensional materials together provides a further engineering opportunity – van der Waals interface symmetry engineering. A prime example is that of WSe$_2$/black phosphorous hetero-interfaces~\cite{akamatsu2021van}: WSe$_2$ (Black Phosphorus) possesses three-fold (two-fold) rotational symmetry. Their interface loses all rotational symmetry allowing a large quantum geometrical current to manifest uniformly in the bulk of the sample even when the individual constituents (WSe$_2$ and black phosphorous) did not exhibit bulk photocurrents~\cite{akamatsu2021van}. This is a striking manifestation of symmetry engineering and van der Waals controlled photocurrent.\\

Such engineering capability can be pushed to the extreme in moir\'e materials where the reconstructed moir\'e minibands can host pronounced quantum geometrical quantities~\cite{liu2020anomalous, kaplan2022twisted,chaudhary2022shift,ma2022intelligent,arora2021strain} that depend not only on the twist angle between individual layers, but also device conditions such as applied gate voltage~\cite{otteneder2020terahertz,ma2022intelligent}. As such, moir\'e minibands can produce a complex structure of photocurrent as a function of polarization/helicity and gate voltages (Fig.~\ref{fig_geometry}\textbf{g-h}). This sensitive dependence on the details of the moire minibands can be exploited as a means of polarimetry for light; it can even enable to identify its wavelength and power all in a single moir\'e device.~\cite{ma2022intelligent} Looking forward, photocurrent sensitivity to symmetry and wavefunctions may render it capable of diagnosing the menagerie of correlated and spontaneously broken symmetry phases in moir\'e materials~\cite{balents2020superconductivity}.\\

{\bf Magnetic photocurrents --} Beyond spatial symmetries, photocurrent diagnostics can be applied to detect quantum properties under broken time-reversal symmetry. Indeed broken time-reversal symmetry (in addition to broken inversion) lead to new classes of magnetic photocurrents (see Box 2 table) that are directly proportional to the magnetic order: when the magnetic order flips, the magnetic photocurrent will also reverse its direction. Two prominent magnetic photocurrents are the circular shift current (shift current induced by circularly polarized light) and linear injection current (injection current induced by linearly polarized light).~\cite{zhang2019switchable,wang2020electrically,ahn2020low} One special class of magnetic materials that have an urgent need for photocurrent characterization are PT symmetric antiferromagnets. PT symmetry guarantees a vanishing net magnetization (full compensation) making the detection of its antiferromagnetic order a challenging, yet highly desirable, task for spintronic applications.~\cite{jungwirth2016antiferromagnetic, vsmejkal2018topological} Magnetic photocurrents can provide a non-invasive measure of the antiferromagnetic order; when combined with scanning techniques (see section 5), PC may even enable spatial resolution of magnetic domains.\\

While there are a steadily increasing range of PT symmetric antiferromagnetic materials~\cite{tang2016dirac}, two-dimensional vdW magnets are a particularly simple and illustrative example. For instance, by combining stacking even layers of the two-dimensional ferromagnet CrI$_3$ together, spins between adjacent layers become aligned antiferromagnetically realizing a layered PT symmetric antiferromagnetic ground state.~\cite{huang2017layer} In a similar fashion, even layered MnBi$_2$Te$_4$ also realizes a layered PT-symmetric antiferromagnet. Even as both have been predicted support magnetic photocurrents~\cite{zhang2019switchable,wang2020electrically}, MnBi$_2$Te$_4$ stands out due to its nontrivial topology that supports Axion electrodynamics.~\cite{wilczek1987two}

\section*{Ultrafast photocurrents: quantum kinetics and out-of-equilibrium material physics}

The quantum kinetics of out-of-equilibrium electronic states can range across a wide multitude of timescales from the femtosecond to picosecond and even nanosecond range. The energy relaxation of a photoexcited electronic system typically proceeds in three broad stages. The first and fastest step is the thermalization of the electron gas by carrier-carrier scattering. In this step, the electrons distribute their energy amongst each other and can establish an out-of-equilibrium hot electron distribution that retains the energy of photo-excited carriers. The second stage involves the cooling of the hot electrons, that is, the transfer of the energy out of the electron gas to the lattice. Finally, once carriers are cooled and reach the band edge, the remaining energy is lost through recombination of electron-hole pairs.\\ 

Ultrafast photocurrent techniques can be used to probe the quantum kinetics in each of these timescales: each timeframe reveals different particle interactions/scattering processes depending on the exact mechanism and material properties of the system. Ultrafast photocurrent techniques combine traditional optical pump-probe~\cite{fischer2016invited} with a variety of electrical read-out schemes
that work together to overcome fundamental electronic bandwidth limitations. In contrast to all-optical techniques, ultrafast photocurrent measurements selectively isolate the electronic system and its associated electronic processes. This allows different (electronic) carrier dynamics and relaxation pathways to be studied depending on the specific microscopic mechanisms underlying the specific photocurrent generation. For instance, the time-evolution of electronic temperature can be probed by photothermoelectric currents, as described below. Similar to photocurrent spectroscopy described above, one key advantage of ultrafast photocurrents compared to all-optical pump-probe measurement is its enhanced signal-to-noise quality when applied to small nanoscale materials whose absorption coefficients are weak.~\cite{sun2012ultrafast,graham2013photocurrent}\\

\begin{figure}[ht]
\centering
\includegraphics[width=\linewidth]{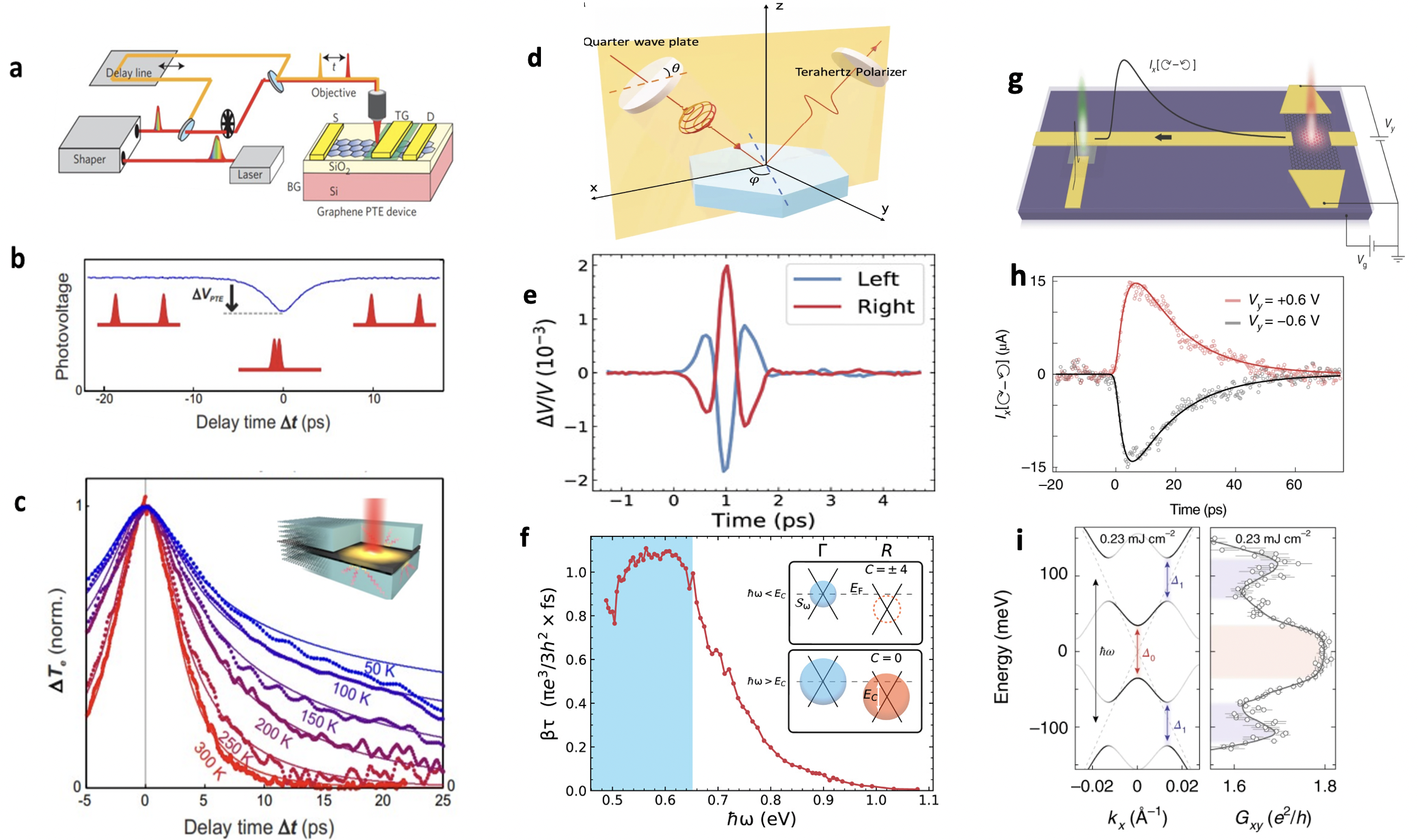}
\caption{\textbf{Ultrafast photocurrents: photocurrent autocorrelation and THz time-domain spectroscopy.} (a) Schematic setup of measuring the autocorrelation of graphene PTE current. (b) Graph plots the generated photovoltage as a function of the time delay $\Delta t$ between two pulses. (c) Electron temperature measurements as a function of the time delay tracks the cooling dynamics of the hot electron gas. Inset illustrates the mechanism for carrier cooling – phonon polariton emission. (d) Schematic of the free-space THz emission as a measure of photocurrent in the time domain. (e) Generated terahertz pulses measured from left and right circularly polarized pump light at 45° angle of incidence in a topological chiral semimetal RhSi~\cite{rees2020helicity}. Their difference is the photon helicity–dependent circular photocurrent signal. (f) The photocurrent amplitude as a function of the excitation photon energy. (g) Schematic of THz time-domain photocurrent spectroscopy. (h) Measurement of the circular photocurrent transient as a function of real time measured in Graphene. (i), Light induced gap openings in Graphene under intense mid-infrared illumination. Panels a and b are adapted from Ref.~\citenum{tielrooij2015generation}. Panel c is adapted from Ref.~\citenum{tielrooij2018out}. Panel d is adapted from Ref.~\citenum{ni2021giant}. Panels e and f are adapted from Ref.~\citenum{rees2020helicity}. Panels g-i are adapted from Ref.~\citenum{mciver2020light}.}
\label{fig_ultrafast}
\end{figure}

{\bf Ultrafast photocurrent autocorrelation -- } In photocurrent autocorrelation measurements, the steady-state photocurrent is measured under excitation by two time-separated optical pulses. In materials wherein photocurrent is a nonlinear function of intensity, the measurement depends on the pulse separation and allows the photocurrent response time to be inferred. Its operating principle can be understood by considering the total time-integrated photocurrent response produced by two pulsed excitations of equal power P separated by a time $\Delta t$, (Fig.~\ref{fig_ultrafast}\textbf{a}). For a sublinear photocurrent response, two pulses coinciding in time $\Delta t = 0$ produce a photocurrent that is less than two pulses that are well separated $\Delta t \to \infty$. Hence, the photocurrent response exhibits a dip around $\Delta t = 0$ (Fig.~\ref{fig_ultrafast}\textbf{b})and saturates as $\Delta t \to \infty$. The saturation point describes the regime where response time is larger than the pulse separation, whereas a small $\Delta t$ describes the intermediate regime where the photocurrent does not have time to completely relax before the second pulse. Therefore, the photocurrent response time can essentially be extracted from the dip in the photocurrent data PC($\Delta t$) (Fig.~\ref{fig_ultrafast}\textbf{b}). This technique has proven powerful in probing the electron dynamics in a range of solid-state systems.~\cite{gabor2012ultrafast, sun2012ultrafast,vogt2020ultrafast}\\

The earliest works and application of the technique involved characterization of semiconductor photoconductive switches~\cite{smith1989picosecond,downey1984picosecond} in which the entire photocurrent lifecycle occurs on sub-picosecond timescales due to fast carrier cooling and short recombination lifetimes caused by carrier trapping via defects. In low-dimensional systems however, energy relaxation pathways are strongly attenuated due to a change in the phase space of quantum kinetic processes. In graphene, for example, slow electron-phonon interactions~\cite{bistritzer2009electronic, song2011hot, song2012disorder} extend the lifetime of the hot-carrier distribution into the picosecond range within the resolution of ultrafast techniques. Because the photocurrent response in graphene is dominated by its hot-electron distribution (established quickly by fast carrier-carrier thermalization~\cite{winzer2010carrier, song2013photoexcited, tielrooij2013photoexcitation, brida2013ultrafast}) and the PTE, photocurrent autocorrelation measurements can directly track the cooling dynamics of the hot-electron gas (Fig.~\ref{fig_ultrafast}\textbf{b-c}). In this regard, photocurrent autocorrelation has been used to probe not only the electron-phonon interactions involved in carrier cooling~\cite{sun2012ultrafast, urich2011intrinsic} but even the initial electron-electron scattering induced thermalization of the hot-electron distribution on femtosecond timescales.~\cite{tielrooij2015generation} These techniques have even revealed unconventional scattering such as supercollision cooling~\cite{graham2013photocurrent} and near-field radiative cooling via phonon polariton modes in remote dielectric layers (Fig.~\ref{fig_ultrafast}\textbf{c}).~\cite{tielrooij2018out}\\

Although lattice cooling occurs much faster in semiconductors, the presence of a gap inhibits rapid recombination, in the absence of defects with large capture cross-sections, so that carriers can reside for some time at the conduction and valence band edge before recombining. Photocurrent response, in these systems, are instead governed by the diffusion of photo-excited carriers in the presence of electric fields and probes interactions that influence this process. For example, the response time is sensitive to the formation of excitons~\cite{wang2018colloquium}
and their many-body physics which can be probed using photocurrent autocorrelation~\cite{wang2015ultrafast, massicotte2018dissociation, massicotte2016picosecond}.\\

{\bf THz time-domain photocurrent spectroscopy -- } THz time-domain photocurrent spectroscopy offers further functionality over the photocurrent autocorrelation technique in that the photocurrent pulse itself is sampled over ultrafast timescales using electro-optic-based detectors such as ZnTe~\cite{dexheimer2017terahertz}. Electro-optic detectors offer the advantage of directly detecting the electric field instead of the intensity (e.g., in absorptive photodetectors). This technique is routinely used in a variety of settings (e.g., bulk quantum materials samples) where an ultrafast pump laser generates transient photocurrents in the crystal, which in turn emits THz radiation that are subsequently detected, Fig.~\ref{fig_ultrafast}\textbf{d},{e}. They are often operated using optical pumps in the visible enabling to study a vast variety of ultra-fast scattering processes and phenomena.~\cite{dexheimer2017terahertz}\\ 

More recently, by tuning the pump photon energy into the infrared range and studying photocurrent as a function of photon energy, near quantized plateaus of circular injection current (Fig.~\ref{fig_ultrafast}\textbf{f}) have been directly observed~\cite{rees2020helicity} -- a ``spectroscopic'' signature of the Chern flux of chiral Weyl semimetals.~\cite{de2017quantized} From a broader perspective, such (variable pump-spectroscopy) techniques naturally lend themselves in interrogating processes that depend on extremely far out-of-equilibrium photo-excited distributions (as opposed to hot but thermalized distributions) such as ultra-fast spin currents in magnetic heterostructures.~\cite{kampfrath2013terahertz, cheng2019far}\\

{\bf On-chip THz detection -- } For nanomaterials of weak emission coefficient, the THz time-domain spectroscopy can be carried out using an {\it on-chip} photoconductive switch. Fig.~\ref{fig_ultrafast}\textbf{g} displays a typical schematic of the measurement set-up. The photoactive device is contacted between two transmission lines that act as waveguides for the transient photocurrent pulse produced by excitation. At some distance down the transmission lines, a probe pulse is used to open a photoconductive switch that samples the photocurrent on a timescale dictated by the response time of the switch. The photocurrent is measured as a function of the delay time between the pump and probe and directly tracks the magnitude of the photocurrent transient in time. In this manner, the technique allows to study the quantum transport of photo-excited carriers~\cite{prechtel2012time, gallagher2019quantum} and their out-of-equilibrium phases.~\cite{prechtel2011time}\\

Another striking attribute of ultrafast photocurrent techniques are that they enable to excite electronic systems with intensities far higher than sustainable in conventional devices under continuous wave illumination. As such they provide access to extreme out-of-equilibrium physics that rely on strong instantaneous electric fields. A particularly striking example of this is Floquet band engineering in graphene~\cite{mciver2020light} to produce unconventional Floquet driven topological phases. By driving graphene with circularly polarized light, theoretical proposals~\cite{Oka2009,kitagawa2011transport} had long predicted an anomalous Hall phase; the requirements on electric field to achieve sizable anomalous Hall effect were challenging: laser intensity requirements exceed those expected to destroy a typical graphene sample under continuous wave irradiation~\cite{mciver2020light}. Ref. ~\citenum{mciver2020light} instead circumvented this by employing an ultrafast time-domain photocurrent technique. They achieved large electric fields required for a sizeable anomalous Hall effect in graphene, but only for a short amount of time, avoiding damage to the graphene device; here the Austin switch enabled to directly measure the (Floquet) Hall conductivity (Figs.~\ref{fig_ultrafast}\textbf{g-i}).\\ 

\section*{Outlook} Looking forward, beyond the specific diagnostics areas we describe above (see Table 1 for a selective summary of key photocurrent diagnostics), photocurrent can also play important roles in research frontiers such as probing correlated electron systems, optical/dynamical control, and their intersection. For example, the PTE current can be used as a probe of correlated insulating gaps in novel correlated systems such as moir\'e materials~\cite{balents2020superconductivity}, because the Seebeck coefficient is highly sensitive to a resistive state; in contrast, conventional transport probes rely on charge current that can easily be shunted along highly conductive regions. Another emerging diagnostic is how geometric photocurrents can be used to probe symmetry breaking and phase transitions due to its high sensitivity to the symmetry of electronic systems. In particular, the energy of mid- and far-infrared photons often directly match the energy scale of correlated quantum phases, such as charge density waves and Mott gaps, which can lead to significantly increased sensitivity. For instance, recent work used photocurrent as a probe of the unique gyrotropic electronic order in a correlated semimetal TiSe$_2$.~\cite{Xu2020TiSe2} From a dynamical control perspective, ultrafast photocurrents are essentially a light-induced transient electric field; such transient electric fields can be used to control the symmetry of electronic states on the ultrafast time scale.~\cite{sirica2022photocurrent} Lastly, by combining the ultrafast optics, polarization control and near-field spatial resolution, next generation photocurrent setups will be able to probe and control the novel physics at the domain walls or edges of novel 2D materials all in-situ. Used in combination, photocurrent techniques are a powerful multi-scale tool that can interrogate a wide range of novel topological, geometrical and correlated physics.

\bibliography{new}

\newpage

\section*{Acknowledgements}
We thank Zihan Wang and Zumeng Huang from Ma lab for COMSOL simulations and figure assistance. Q.M. was supported through NSF Career DMR-2143426 and the CIFAR Azrieli Global Scholars program. R.K.K. acknowledges the EU Horizon 2020 program under the MarieSkłodowska-Curie grants 754510 and 893030. S.-Y.X. was supported through NSF Career (Harvard fund 129522) DMR-2143177. F.H.L.K. acknowledges  support from the ERC TOPONANOP  (726001), the government of Spain (PID2019-106875GB-I00; Severo Ochoa CEX2019-000910-S [MCIN/ AEI/10.13039/501100011033]), Fundació Cellex, Fundació Mir-Puig, and Generalitat de Catalunya (CERCA, AGAUR, SGR 1656). Furthermore, the research leading to these results has received funding from the European Union’s Horizon 2020  under grant agreement  no. 881603 (Graphene flagship Core3) and 820378 (Quantum flagship). J.C.W.S acknowledges support from the Singapore MOE Academic Research Fund Tier 3 Grant MOE2018-T3-1-002. 

\newpage

\noindent\fbox{
\parbox{\textwidth}{
\section*{Box 1: Long-range photocurrent and remote sensing}

Spatially localized photocurrent generation and the remote contact sensing in conductors can be conveniently captured by a Shockley-Ramo-type approach~\cite{song2014shockley} that has been successfully applied to describe the details of a variety of scanning photocurrent spatial maps in quantum materials~\cite{hesp2021nano, sunku2021hyperbolic,gabor2011hot,wang2019robust, ma2019giant, cao2016photo, woessner2016near, seifert2019quantized, shao2021nonlinear, mayes2020current}. In these, a spatially localized photoexcitation (e.g., a tightly focused laser spot) generates a local photocurrent ${\bf j}_{\rm loc}({\bf r})$. Even as the localized photoexcitation can be spatially far away from current collecting contacts, ${\bf j}_{\rm loc}({\bf r})$ 
sets up a local electromotive force that can drive ambient carriers outside of the excitation region and into the remote contacts. Current collected at the remote contacts, $I$, follow the simple form~\cite{song2014shockley} 
\begin{equation}
    I = A \int {\bf j}_{\rm loc}({\bf r})  \cdot \nabla \psi({\bf r})  d{\bf r}
    \label{eq:SR}
\end{equation}
where $A$ is a prefactor which depends on device and external circuit configuration, and $\psi({\bf r})$ is a weighting field obtained by solving a Laplace problem $\nabla \sigma^T ({\bf r}) \nabla \psi =0$ in the device geometry and contact configuration where photocurrent is measured. Here $\sigma ({\bf r})$ is the local conductivity tensor. The general form of Eq.~(\ref{eq:SR}) is ubiquitous appearing for instance in the original Shockley-Ramo theorem~\cite{shockley1938currents,ramo1939currents} traditionally used in the context of vacuum-tube electronics, but also in descriptions of photo-responsivity in highly inhomogeneous devices~\cite{hesp2021nano, lundeberg2020thermodynamic}. 
\\

Importantly, $\nabla \psi({\bf r})$ describe ``Shockley-Ramo'' streamlines along which diffusion currents prefer to flow and enables $I$ to sense the direction in which ${\bf j}_{\rm loc}({\bf r})$ flows. In a simple two-terminal rectangular geometry, such streamlines flow uniformly from one contact into the ground (panel A and B). As a result, local photocurrent generated far away from contacts can be sensed remotely by $I$, with a sign determined by whether local photocurrent ${\bf j}_{\rm loc}({\bf r})$ is aligned or anti-aligned with the streamline pattern $\nabla \psi({\bf r})$: e.g., local photocurrent generated in the bulk enable to identify p-n interfaces (panel A); similarly, when local photocurrent is generated along edges [e.g., in topological materials, see main text], it also can be detected remotely (panel B). Note the strength of $I$ depends on the density of the streamlines allowing strong signals to be collected even from photocurrent generated far from contacts. Streamlines (and hence the responsivity) can be contorted by the device geometry (Panel C), by floating (non-current collecting) contacts (Panel D), and as well as by toggling contact configuration that can be used to infer local photocurrent direction. 

\includegraphics[width=0.9\linewidth]{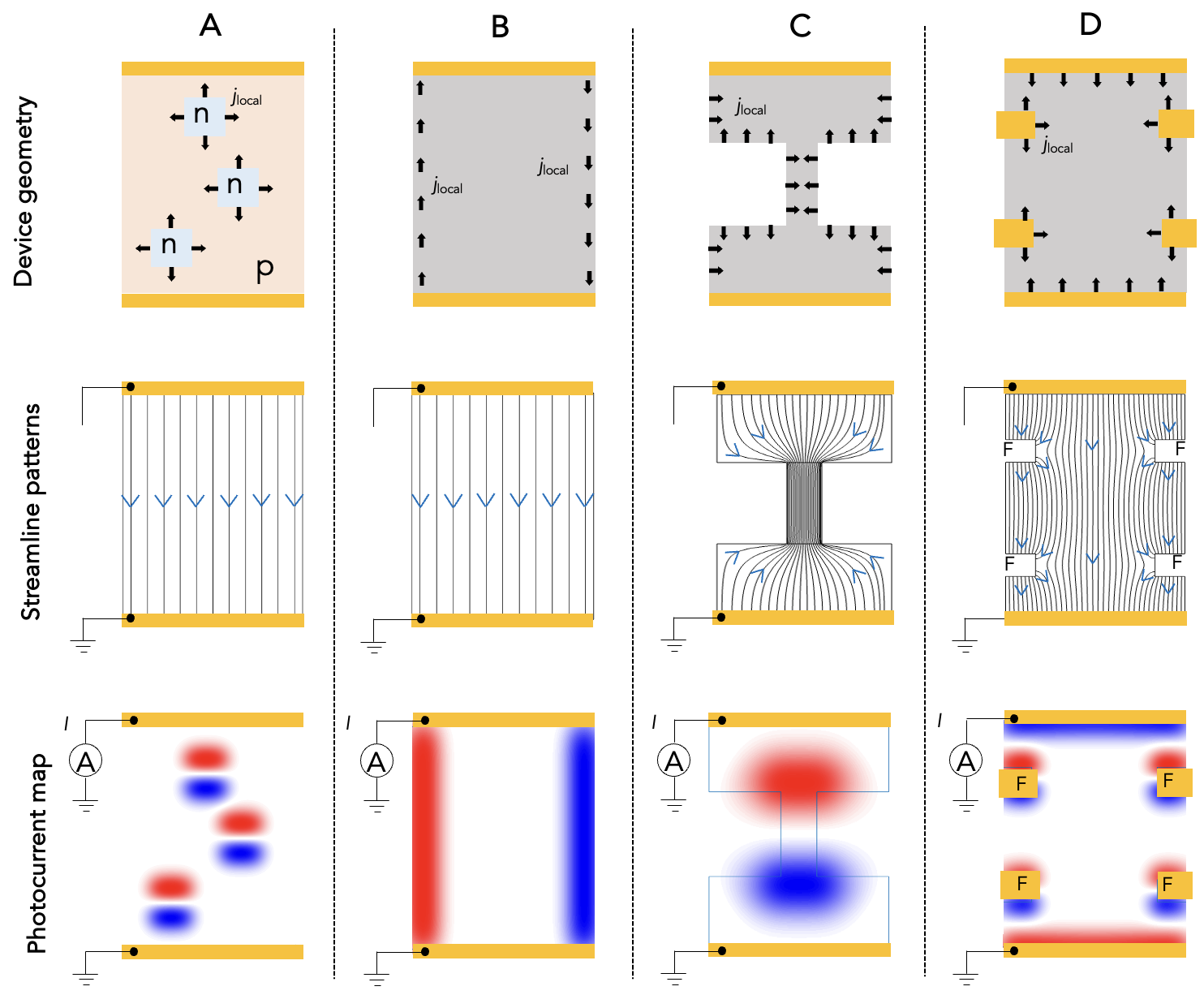}
}}

\noindent\fbox{
\parbox{\textwidth}{
\section*{Box 2: Types of quantum geometric photocurrents and symmetry}

Two prominent quantum geometric photocurrents that arise from interband transitions (from an initial to a final state) are injection and shift photocurrents.~\cite{Sipe00,ahn2020low} The injection current arises from the change in carrier group velocity as electrons are photoexcited from an initial to a final state. The shift current, on the other hand, manifests from a real-space shift as electrons are photoexcited.~\cite{Morimoto:2016iu} It is often expressed as a shift vector~\cite{von1981theory,Sipe00} that depends on the polarization of light and captures the real-space displacement when an electron is excited; the shift vector can trace its origins to the geometric Pancharatnam-Berry phase accrued during the transition.~\cite{xiong2021atomic, wang2022generalized}\\

A further recurrent delineation (applied broadly across photocurrent mechanisms) is that of photocurrent responses under specific light polarization conditions, namely circular or linearly polarized light. This distinction is useful since circular (helicity-dependent) photocurrents and linearly photocurrents often generate contrasting behavior, and, different material symmetry conditions are either forbidden/allowed (see table below).\\

A striking example are non-magnetic materials wherein only circular injection and the linear shift quantum geometric photocurrents are allowed. The circular injection photocurrent can be understood as follows: the combination of inversion symmetry (P-symmetry) breaking and time-reversal symmetry (T-symmetry) gives rise to opposite Berry curvature distributions at momenta $k$ and $-k$, see illustration (red vs blue). These produce optical selection rules that lead to an imbalanced photoexcitation rate for electrons at $k$ and $-k$ under circularly polarized light. The non-uniform k-space photoexcited carrier distribution that ensues possess carriers with uncompensated group velocity and, therefore, non-zero net current.\\

The linear shift photocurrent can be understood in an analogous way. P symmetry ensures that the shift vector, characterizing the real space shift between a photoexcited final state and initial state, is an odd function: shift vector at $k$ has an opposite sign from shift vector at -$k$. When P-symmetry is broken however, this condition is removed allowing optical excitation to induce a shift of the charge carrier’s real space position and a charge current, see wavepacket in illustration. Interestingly, in materials with low enough spatial symmetry, the contribution from different linear polarizations do not compensate with each other enabling non-zero shift photocurrent even under unpolarized light.~\cite{cook2017design}\\

In magnetic materials (broken T symmetry), new magnetic photocurrents can emerge. They include the circular shift photocurrent (shifted wavepackets have opposite signs for different helicity: red vs blue wavepackets in illustration) and linear injection photocurrents (excited carriers have uncompensated group velocity). In a special class of magnets, e.g., PT symmetric antiferromagnets, a composite PT symmetry prohibits the nonmagnetic photocurrents discussed above, allowing to isolate magnetic photocurrents.~\cite{wang2020electrically,ahn2020low,watanabe2021chiral}
\\\\
\includegraphics[width=\linewidth]{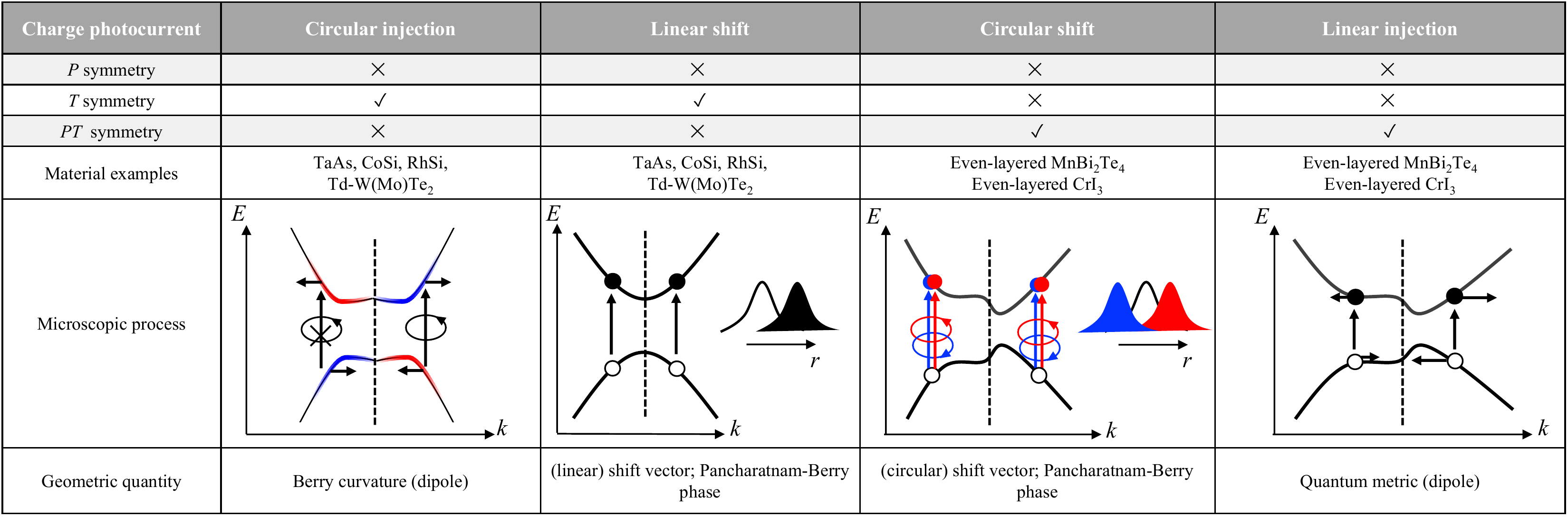}
}}

\newpage

\newpage

\newpage

\newpage

\newpage

\newpage

\setcounter{figure}{0}
\renewcommand{\figurename}{Table}

\begin{figure}[ht]
\centering
\includegraphics[scale=0.57]{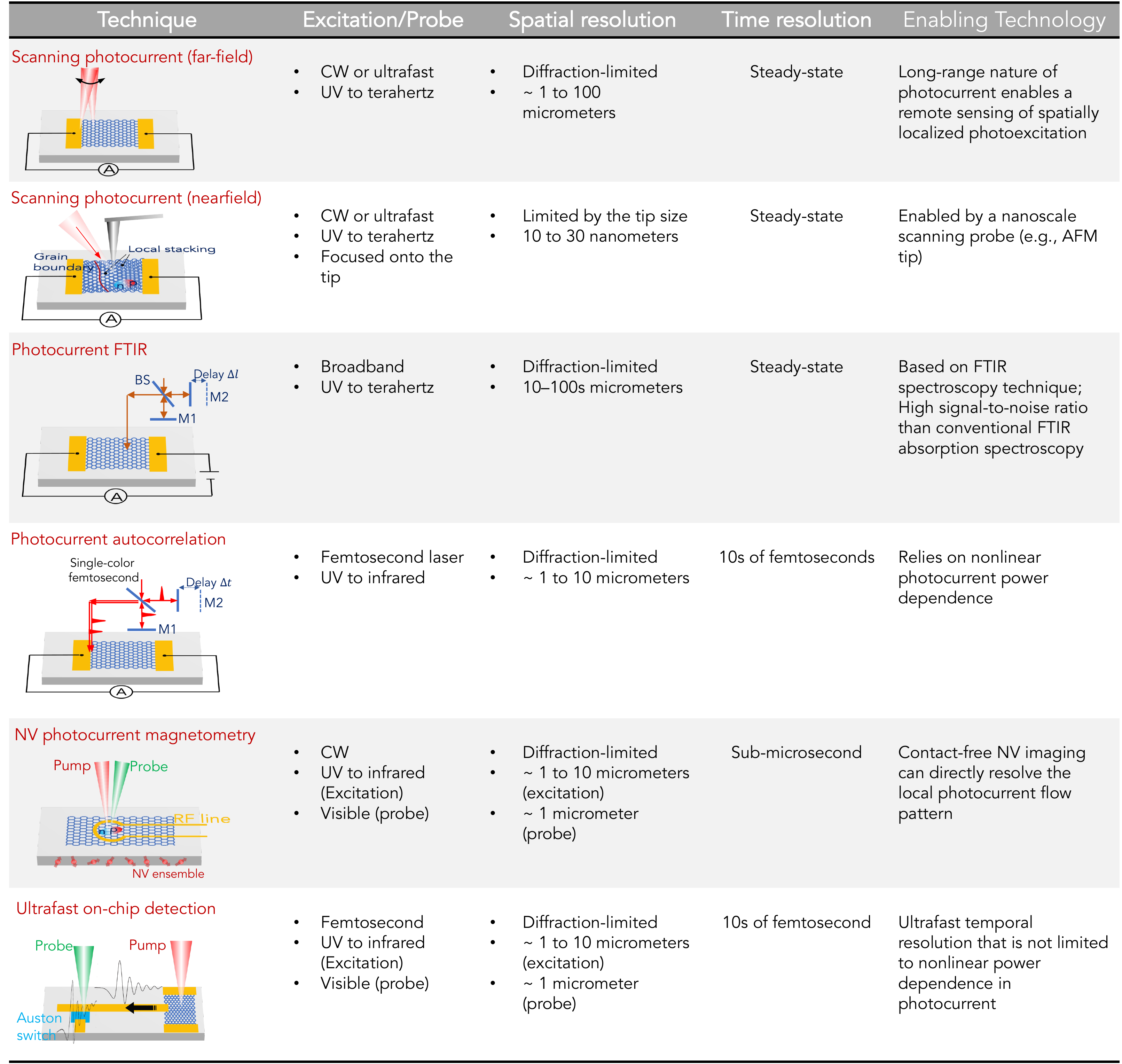}
\caption{{\bf Powerful photocurrent diagnostics of spatiotemporal characters enabled by various imaging and spectroscopy techniques}.} 
\label{fig_technique}
\end{figure}

\newpage

\end{document}